# Use of a speed equation
# for numerical simulation of hydraulic fractures

Alexander M. Linkov

*Institute for Problems of Mechanical Engineering, 61, Bol'shoy pr. V. O., Saint Petersburg, 199178, Russia*
*Presently: Rzeszow University of Technology, ul. Powstancow Warszawy 8, Rzeszow, 35-959, Poland,*
*e-mail: linkoval@prz.edu.pl*

***Abstract.*** This paper treats the propagation of a hydraulically driven crack. We explicitly write the local speed equation, which facilitates using the theory of propagating interfaces. It is shown that when neglecting the lag between the liquid front and the crack tip, the lubrication PDE yields that a solution satisfies the speed equation identically. This implies that for zero or small lag, the boundary value problem appears ill-posed when solved numerically. We suggest $\varepsilon$- regularization, which consists in employing the speed equation together with a prescribed BC on the front to obtain a new BC formulated at a small distance $\varepsilon$ behind the front rather than on the front itself. It is shown that $\varepsilon$- regularization provides accurate and stable results with reasonable time expense. It is also shown that the speed equation gives a key to proper choice of unknown functions when solving a hydraulic fracture problem numerically.

*Keywords:* hydraulic fracturing, numerical simulation, speed equation, ill-posed problem, regularization

## 1. Introduction

Hydraulic fracturing is a technique used extensively to increase the surface to or from which a fluid flows in a rock mass. It is applied for various engineering purposes such as stimulation of oil and gas reservoir recovery, increasing heat production of geothermal reservoirs, measurement of in-situ stresses, control of caving in the roof of coal and ore excavations, enhancing efficiency of $CO_2$ sequestration and isolation of toxic substances in rocks. In natural conditions, a similar process occurs when a pressurized melted substance fractures impermeable rock leading to the formation of veins of mineral deposits. Beginning with researchers such as Khristianovich & Zheltov (1955), Carter (1957), Perkins & Kern (1961), Geertsma & de Klerk (1969), Howard & Fast (1970), Nordgren (1970), Spence &Sharp (1985), Nolte (1988) numerous studies have been published on the theory and numerical simulation of hydraulic fracturing (see, e. g., Desroches et al. 1994; Lenoach 1995; Garagash & Detournay 2000; Adachi & Detournay 2002; Detournay et.al. 2002; Savitski & Detournay 2002; Jamamoto et al. 2004; Pierce & Siebrits 2005; Garagash 2006; Adachi et al. 2007; Mitchel et al. 2007; Kovalyshen & Detournay 2009; Kovalyshen 2010; Hu & Garagash 2010; Garagash et al. 2011 and detailed reviews in many of them). The review by Adachi et al. (2007) is specially organized to give a comprehensive report on the computer simulation; it actually covers the present state of the art as well. Thus there is no need to dwell on the historical point, detailed analysis of the processes in a near-tip zone resulting in particular asymptotics and regimes of flow, general equations and general approaches used to date. Being interested in computer aided simulation of hydraulic fracturing, in this paper we would rather focus on a key area that needs to be addressed for further progress in *numerical simulation* (Adachi et al. 2007, p. 754); this is the need "to dramatically speed up" simulators of fracture propagation. Naturally, reasonable accuracy of results should be guaranteed. The goal cannot be reached without clear understanding of underlying *computational difficulties* which strongly influence the accuracy and stability of numerical results and robustness of procedures.

The paper addresses this issue. It presents in detail the results of brief communications by the author (Linkov 2011a, b). Our prime objective is to delineate and to overcome the computational difficulty caused, in essence, by strong non-linearity of the lubrication equation and by the moving boundary. In contrast with the cited previous publications which employed the *global* form of the mass balance to trace liquid front propagation in time, we explicitly write and use the *local* speed equation (SE). The speed function entering the SE may serve to employ methods developed in the theory of propagating surfaces (Sethian 1999). The SE gives also a key to proper choice of unknown functions, which are analytical up to the front. We show that at points where the lag between the liquid front and the crack tip is zero, the lubrication equation yields that its solution identically satisfies the SE. This implies that for zero or small lag, the problem will appear ill-posed when solved numerically for a fixed front at a time step, and consequently it requires appropriate regularization to have accurate and stable numerical results. We suggest a method of regularization that employs the very source of the difficulty to overcome it. The SE and a BC at the front are used together to derive a new BC formulated at a small distance $\eta_\varepsilon$ behind the front rather than on the front itself. This leads to $\varepsilon$-regularization which provides accurate and stable numerical results. The Nordgren model serves to clearly display the computational features discussed. It is also used to obtain benchmarks with five correct significant digits, at least.

## 2. Global mass balance, speed equation and ill-posed problem for hydraulically driven fracture

Reynolds transport theorem (e. g. Crowe et al. 2009), applied to the mass of an arbitrary volume of a medium in a narrow channel between closely located boundaries, after averaging over the channel width, reads:

$$\frac{dM}{dt} = \int_{S_m} \left( \frac{\partial \rho}{\partial t} w + \rho \frac{\partial w}{\partial t} \right) dS + \int_L \rho w v_n dL , \qquad (2.1)$$

where $dM/dt$ is the external mass coming into or out of the considered volume per unit time, $S_m$ is the middle surface of the volume, $w$ is the width (opening) of the channel, $L$ is its contour, $\rho$ is the mass density averaged over the width, $v_n$ is the normal to $L$ component of the particle velocity also averaged over the width; the normal $\boldsymbol{n}$ is assumed to be in the plane tangent to $S$ at a considered point. Applying (2.1) to the total volume of the medium with the middle surface $S_t$ and contour $L_t$ at a moment $t$ we have:

$$\frac{dM_e}{dt} = \int_{S_t} \left( \frac{\partial \rho}{\partial t} + \rho \frac{\partial w}{\partial t} \right) dS + \int_{L_t} \rho_* w_* v_{n*} dL . \qquad (2.2)$$

Here, and henceforth, the star denotes that a value refers to the medium front. When writing (2.2) we assume that there is no significant sucking or vaporization through the front. Then the speed $V_*$ of the front propagation coincides with the normal component $v_{n*}$ of the particle velocity. Thus we have the key equation:

$$V_* = v_{n*} = \frac{dx_{n*}}{dt} , \qquad (2.3)$$

where $x_{n*}$ is the normal to the front component of a position vector of a point on the front; in global coordinates, $\mathbf{x} = \mathbf{x}_*(t)$ is a parametric equation of the front contour $L_t$. As a rule, the tangent component of the particle velocity is small as compared with the normal component at the front. In this case, the front moves with the speed exactly equal to the velocity of fluid particles comprising it, and we have $V_* = v_{n*} = |\mathbf{v}_*|$, where $\mathbf{v}_*$ is the vector of the particle velocity averaged over the front width.

For incompressible homogeneous liquid, $\rho = const$, and (2.2) may be written as the equation of the liquid volume balance:

$$\frac{dV_e}{dt} = \int_{S_t} \frac{\partial w}{\partial t} dS + \int_{L_t} w_* v_{n*} dL , \qquad (2.4)$$

where $V_e = M_e / \rho$ .

*Comment 1.* For 1-D case, the volume conservation equation (2.4) yields $\int_0^{x_*(t)} w(x,t)dx = V_e(t)$. Then for a 'rigid-wall' channel ($w = w(x)$, $\partial w/\partial t = 0$), integration gives the front location as a function of time $x_*(t) = f(V_e(t))$. It is easily seen from the last expression that if the width $w(x)$ decreases fast enough with growing $x$, then for a prescribed influx $V_e(t)$, the front coordinate turns to infinity at a finite time $t_*$ ($x_*(t_*) = \infty$). A solution does not exist for $t > t_*$. For instance, in the case of the constant influx rate $q_0$ ($V_e(t) = q_0 t$) and exponentially decreasing width ($w = a\exp(-\alpha x)$, $a$, $\alpha > 0$), the front turns to infinity as $t \to t_* = a/(\alpha q_0)$. There is no solution for $t > t_*$. This clearly indicates that problems involving flow of *incompressible* liquid in a thin channel are quite tricky. A solution may not exist or it might be difficult to find the solution numerically, especially when the rigidity of channel walls is very high (according to Pierce & Siebrits (2005), high rigidity leads to a stiff system of ordinary differential equations, when solving a boundary value problem by finite differences).

By definition of the flux through the channel width (e.g. Batchelor 1967), we have $\mathbf{q} = w\mathbf{v}$. Then in (2.4) $w_*v_{n*} = q_{n*}$ is the normal to $L$ component of the flux through the channel width at the liquid front, and equation (2.4) may be written as

$$V_* = v_{n*}(\mathbf{x}_*) = \frac{q_{n*}(\mathbf{x}_*)}{w_*(\mathbf{x}_*)}. \tag{2.5}$$

This is the speed equation (SE). Its right hand side (r. h. s) defines the so-called speed function. Emphasize that the SE is general. It is not influenced by viscous properties of a particular incompressible liquid and by the presence of leak-off or influxes through the walls of a thin channel.

In some cases, for instance, when a liquid flows in a fracture without a lag between the liquid front and the crack tip, both the flux and opening turn to zero at the liquid front. Then (2.5) takes the limit form:

$$v_{n*}(\mathbf{x}_*) = \lim_{\mathbf{x} \to \mathbf{x}_*} \frac{q_n(\mathbf{x})}{w(\mathbf{x})}. \tag{2.6}$$

It is highly apparent that even when $w_*(\mathbf{x}_*) = 0$ and $q_{n*}(\mathbf{x}_*) = 0$, the limit on the r. h. s. of (2.6) should be finite to exclude the front propagation with infinite velocity. This suggests using the particle velocity as an unknown in numerical calculations, because it is non-singular in entire flow region including the front. What is also beneficial, the velocity is non-zero except for flows with return points.

In formulations of problems for hydraulic fracturing, the average particle velocity $\mathbf{v}(\mathbf{x}) = \mathbf{q}(\mathbf{x})/w(\mathbf{x})$ does not enter equations. Rather, the total flux $\mathbf{q}$ through a cross-section is used. Specifically, the divergence theorem applied to (2.1) in the case of incompressible liquid yields the continuity equation in terms of the flux $\mathbf{q}$ and opening:

$$\frac{\partial w}{\partial t} + div\mathbf{q} = q_e, \tag{2.7}$$

where $q_e$ is the prescribed intensity of distributed external sources of liquid, $\mathbf{q}$ is the flux vector, defined in the tangent plane to $S_t$ at a considered point. The Poiseuille law, used for a flow of incompressible liquid in a narrow channel, connects the flux $\mathbf{q}$ with the pressure gradient

$$\mathbf{q} = -D(w, p)\mathrm{grad}p. \tag{2.8}$$

Herein, $D$ is a prescribed function or operator, gradient is also confined to the coordinates in the tangent plane. Substitution of (2.8) into (2.7) gives the lubrication partial differential equation (PDE) at points of liquid:

$$\frac{\partial w}{\partial t} - div \bigcirc (w, p)\mathrm{grad}p - q_e = 0. \tag{2.9}$$

Some initial (normally zero-opening) condition is assumed to account for the presence of the time derivative in (2.9). Being of the second order and elliptic in spatial derivatives, equation (2.9) requires a boundary condition (BC) on the liquid contour $L_l$. Normally it is the condition of the prescribed flux $q_0$ at a part $L_q$ and of the prescribed pressure $p_0$ at the remaining part $L_p$ of the contour $L_l$:

$$q_n(\mathbf{x}_*) = q_0(\mathbf{x}_*) \quad \mathbf{x}_* \in L_q; \qquad p(\mathbf{x}_*) = p_0(\mathbf{x}_*) \quad \mathbf{x}_* \in L_p. \tag{2.10}$$

We see that neither partial differential equations (PDE) (2.7)-(2.9), nor BC (2.10) involve the particle velocity. The latter enters only the SE (2.5) or its limit form (2.6) for points on the front. At these points, it is defined by formula (2.3). Since according to (2.8) $q_n = -D(w,p)\partial p / \partial n$ for any direction $\mathbf{n}$ in a tangent plane, the flux entering the numerator on the r. h. s. of (2.5) is

$$q_{n*} = -D(w,p)\frac{\partial p}{\partial n}\bigg|_{\mathbf{x}=\mathbf{x}_*}. \tag{2.11}$$

Then for the hydraulic fracturing, the SE (2.5) is specified as:

$$v_{n*} = -\frac{1}{w_*(\mathbf{x}_*)}D(w,p)\frac{\partial p}{\partial n}\bigg|_{\mathbf{x}=\mathbf{x}_*}, \tag{2.12}$$

where $\mathbf{n}$ is the outward normal to $L_t$ in the tangent plane at a considered point of the liquid front. The r. h. s. of (2.12) specifies the speed function, which is the basic concept of the theory of propagating interfaces (Sethian 1999). It may serve to employ level set methods and fast marching methods.

In hydraulic fracture problems, the opening is unknown. To have a complete system, the lubrication PDE (2.9) is complemented with an equation of solid mechanics connecting the opening $w$ and pressure $p$:

$$A(w,p) = 0. \tag{2.13}$$

As a rule, the operator $A$ in (2.13) is prescribed by using the theory of linear elasticity. In addition, to let the fracture propagate, we need a fracture criterion (Otherwise, the liquid front reaches the crack contour and stops.). Commonly, at points of the crack contour, authors impose the condition of linear fracture mechanics:

$$K_I = K_{Ic}, \tag{2.14}$$

where $K_I$ is the stress intensity factor (SIF), $K_{Ic}$ its critical value.

At the crack contour $L_c$, the opening is set zero:

$$w(\mathbf{x}_c) = 0. \tag{2.15}$$

In general, for physical reasons, excluding negative pressure, which tends to minus infinity at points of the front, where the latter coincides with the crack contour, the liquid surface $S_t$ is solely a part of the crack surface $S_c$ so that the liquid contour $L_t$ is within the crack contour $L_c$. Thus, in general there is a lag between $L_t$ and $L_c$. In this case, the second of the conditions (2.10) is prescribed at the boundary of the propagating liquid front:

$$p(\mathbf{x}_*) = p_0(\mathbf{x}_*). \tag{2.16}$$

In many cases, the change of the pressure $p_0(\mathbf{x}_*)$ along the liquid front is small. Consequently, the tangential derivative $\partial p / \partial \tau$ of the pressure is small as compared with the normal derivative $\partial p / \partial n$, and (2.8) implies that the tangential component of the flux is small as compared with its normal component. Since $\mathbf{v} = \mathbf{q}/w$, it means that the tangential component of the fluid velocity at the front may be neglected. Then, as mentioned above, the front velocity equals to the particle velocity itself. This case will be under discussion further on. We see that the existence of a lag, whatever small it is, has important physical implications for fracture propagation.

Normally the lag is small and accounting for it strongly complicates a problem, while neglecting it may be justified (e. g., Garagash & Detournay 2000). For these reasons, many papers on hydraulic fracturing (e. g., Spence & Sharp 1985; Lenoach 1995; Garagash & Detournay 2000; Adachi & Detournay 2002; Detournay et. al. 2002; Savitski & Detournay 2002; Jamamoto et al. 2004; Pierce & Siebrits 2005; Adachi et al. 2007; Mitchel et al. 2007; Hu & Garagash 2010; Garagash et al. 2011) assume that the lag is zero. Then at all points of the propagating liquid front, coinciding in this case with the crack contour, the flux is zero: $q_{n*}(\mathbf{x}_*) = 0$. The latter condition is met in view of (2.15), because the operator $D(w, p)$ in (2.11) is such that $D(0, p) = 0$. Still, to satisfy both (2.15) and (2.14), the elasticity equations require specific asymptotic behavior of the opening with the coefficient of the asymptotic proportional to the SIF, when the latter is not zero. The SIF depends on the pressure (see, e. g. Spence & Sharp 1985). As a result, at points of the front we have the boundary condition (2.14) which, similar to (2.16), involves the pressure.

The fact that the lag is small and it is commonly neglected has significant consequences for numerical calculations. To show it, consider the problem in the local Cartesian coordinates $x_1'O_1'x_2'$ with the origin $O_1'$ at a point $\mathbf{x}_*$ at the liquid front and the axis $x_1'$ opposite to the direction of external normal to the front at this point. We employ aforementioned advantages of using the particle velocity $\mathbf{v} = \mathbf{q}/w$. In the local system, for points close to the front, the normal component of the velocity is $v = \frac{1}{w}D(w,p)\frac{\partial p}{\partial x_1'}$. Then the lubrication PDE (2.9) at points near the front takes the form:

$$\frac{\partial v}{\partial x_1'} + (v - v_*)\frac{\partial \ln w}{\partial x_1'} - \frac{\partial \ln w}{\partial t}\bigg|_{x_1'=const} + q_e = 0, \qquad (2.17)$$

where the partial time derivative is evaluated under constant $x_1'$. Using $\ln w$ serves to account for an arbitrary power asymptotic behavior of the opening $w(x_1',t) = C(t)(x_1')^\alpha$ ($\alpha \geq 0$) at the front. In particular, in the case of Newtonian liquid, for zero lag, the exponent $\alpha = 2/3$ when in (2.14) $K_{Ic} = 0$ and $\alpha = \frac{1}{2}$ when $K_{Ic} \neq 0$ (Spence & Sharp 1985). In the case of non-Newtonian liquid, formulae for $\alpha$ are given by Adachi & Detournay (2002). For the Nordgren problem discussed below, $\alpha = 1/3$. Note that for a flow with Carter's leak-off, so-called intermediate asymptotics may appear (e.g., Lenoach 1995; Kovalyshen & Detournay 2010). These asymptotics manifest themselves at some distance from a crack contour rather than at the contour itself. For this reason, we shall not use them below in equations involving a point on the front.

When the opening has the power asymptotic near the front, it is reasonable, in addition to the particle velocity, to use also the variable $y = w^{1/\alpha}$, which is linear in $x_1'$ near the front. The PDE (2.17) then becomes

$$\frac{\partial v}{\partial x_1'} + \alpha\frac{v - v_*}{y}\frac{\partial y}{\partial x_1'} - \alpha\frac{1}{y}\frac{\partial y}{\partial t}\bigg|_{x_1'=const} + q_e = 0. \qquad (2.18)$$

Note that the derivative $\partial v/\partial x_1'$ and, under the supposed asymptotic, the derivatives $\partial y/\partial x_1'$, the multiplier $(v - v_*)/y$ and the term $(1/y)\partial y/\partial t$ in (2.18) are finite at the liquid front. Therefore, *the variables* $\mathbf{v} = -\dfrac{D(w,p)}{w}\,\mathrm{grad}\,p$ *and* $y = w^{1/\alpha}$ *present a proper choice of unknown functions for problems of hydraulic fracturing*.

Write (2.18) as

$$y\frac{\partial v}{\partial x_1'} + \alpha(v - v_*)\frac{\partial y}{\partial x_1'} - \alpha\frac{\partial y}{\partial t}\bigg|_{x_1'=const} + yq_e = 0. \qquad (2.19)$$

Equation (2.19) implies that $v = v_*$ at any point, where the opening $w$, and consequently $y = w^{1/\alpha}$, is zero. Such are points at the front for zero lag. This means that when neglecting the lag, the SE (2.12) is satisfied identically by a solution of the PDE (2.18). Obviously, the same holds for a solution of the starting PDE (2.9).

We see that for zero lag, when solving the boundary value (BV) problem for (2.9), one implicitly has satisfied the SE (2.12) additional to the prescribed BC of zero opening (2.15). Note that for the mentioned power asymptotic of the opening, the SE may be re-written in terms of the normal derivative of $y = w^{1/\alpha}$ as $\left.\dfrac{\partial y}{\partial x'_1}\right|_{x'_1 = 0} = Bv_*$, where $B$ is a function of time only. Then at a point of the front, we have satisfied two equations: $y(x_*) = 0$ and $\left.\dfrac{\partial y}{\partial n}\right|_{x = x_*} = -Bv_*$. Recall that the operator div $\bigcirc(w, p)$grad$\tilde{p}$ is elliptic and requires only *one* BC, whereas actually there are *two* BC. Consequently, we have a Cauchy problem for the elliptic operator. As known (e. g. Lavrent'ev & Savel'ev 1999), such a problem is *ill-posed* in the Hadamard sense (1902). To have accurate and stable results when solving an ill-posed problem numerically, one needs its proper regularization (e. g. Tychonoff 1963; Lavrent'ev & Savel'ev 1999).

In the case when the lag is not neglected, we come to similar conclusions if the lag is small enough. In this case, at a point of the liquid front, we have the BC of (2.16) type. As the lag is small, the opening $w_*(\mathbf{x}_*)$ at the liquid front is small, as well. Then from (2.19) it follows that for a solution of the lubrication PDE, the SE is met approximately. Therefore, at points of the liquid front, in addition to the BC (2.16) for the pressure, we have the approximate equation $\left. -\dfrac{1}{w_*(\mathbf{x}_*)} D(w, p) \dfrac{\partial p}{\partial n}\right|_{\mathbf{x} = \mathbf{x}_*} \approx v_{n*}$ for its normal derivative. Hence, the problem of solving the lubrication PDE will appear ill-posed in numerical calculations having the accuracy less than that of the approximate equation. It is reasonable to have a method of regularization, which removes computational difficulties caused by the discussed feature.

*Comment 2*. If a problem is self-similar and the lag is neglected, then integration of the lubrication equation in the automodel coordinate (e. g. Spence & Sharp 1985; Adachi & Detournay 2002) from the liquid front removes the difficulty: it is sufficient to seek the solution by taking into account for asymptotic representation of the opening and pressure near the liquid front. In papers by Spence & Sharp (1985) and Adachi & Detournay (2002), such representations meet both conditions (2.15) and (2.6), which uniquely define the coefficients of the asymptotics at one point (the crack tip). In fact, the authors solve an initial value (Cauchy) *well-posed* problem. The boundary condition of the prescribed flux at another point (the inlet) is not used: the corresponding influx is found *after* obtaining the solution of the Cauchy problem. Similar approach is applicable to the Nordgren problem. It serves us to obtain the benchmarks in Sec. 4. Unfortunately, this method cannot be applied in a general case when a self-similar formulation is not available.

### 3. Nordgren problem. Evidence of ill-posed problem

Consider the Nordgren (1972) model to see unambiguously that the BV problem is ill posed, to find a proper means for its regularization and to obtain accurate numerical results, which may serve as benchmarks. The analysis also confirms that the front velocity does satisfy the asymptotic equation (2.6) despite that both the opening $w(\mathbf{x}_*)$ and the flux $q_n(\mathbf{x}_*)$ are zero at the front.

#### 3.1. Problem formulation

Recall the assumptions of the Nordgren (1972) problem. Similar to the Perkins-Kern (1961) model, it is assumed that a vertical fracture of a height $h$ (Fig. 1) is in plane-strain conditions in vertical cross sections perpendicular to the fracture plane. The cross section is elliptical and the maximal opening $w$ decreases along the fracture. Nordgren's improvement of the model includes finding the fracture length $x_*(t)$ as a part of the solution. Nordgren also accounts for the fluid loss due to leak-off. The corresponding term is actually a prescribed function of time; we neglect it to not overload the analysis. In this case, the continuity equation (2.7) reads $\partial q / \partial x + \partial w / \partial t = 0$, where $w$ is the average opening in a vertical cross section, $q$ is the flux through a cross section divided by the prescribed height $h$. The liquid is assumed Newtonian with the dynamic viscosity $\mu$. Then in (2.8), $D(w,p) = k_l w^3$, where $k_l = 1/(\pi^2 \mu)$ in the case of an elliptic cross section, considered by Nordgren (1972); for an arbitrary thin plane channel, the Poiseuille value $k_l = 1/(12\mu)$ is often used (e. g., Garagash & Detournay 2000; Savitski & Detournay 2002; Garagash 2006; Mitchel et al. 2007; Hu & Garagash 2010). Thus the equation (2.8) becomes:

$$q = -k_l w^3 \frac{\partial p}{\partial x} \,. \tag{3.1}$$

The dependence (2.13) between the average opening and pressure is taken in the simplest form

$$p = k_r w, \tag{3.2}$$

found from the solution of a plane strain elasticity problem for a crack of the height $h$; $k_r = (2/\pi h)E/(1-\nu^2)$, $E$ is the rock elasticity modulus, $\nu$ is the Poisson's ratio. Therefore, for a non-negative opening, the pressure is non-negative behind the front. With lag neglected, the condition (2.15) in view of (3.2) implies that the pressure becomes zero at the liquid front. The opening (as well as the pressure) should then be positive behind the front and zero ahead of it. Under these assumptions, there is no need in the fracture criterion (2.14).

In view of (3.1) and (3.2), the continuity equation (2.9) becomes the Nordgren PDE:

$$\frac{1}{4} k_l k_r \frac{\partial^2 w^4}{\partial x^2} - \frac{\partial w}{\partial t} = 0 \,. \tag{3.3}$$

It is solved under the initial condition of zero opening

$$w(x,0) = 0 \tag{3.4}$$

for any $x$ along the prospect path of the fracture.

The BC for the partial differential equation (PDF) (3.3) includes the condition of the prescribed influx $q_0$ at the fracture inlet $x = 0$:

$$q(0,t) = q_0 \tag{3.5}$$

and the condition that there is no lag between the crack tip and the liquid front:

$$w(x_*, t) = 0. \tag{3.6}$$

The solution should be such that the opening is positive behind the front and zero ahead of it:

$$w(x,t) > 0 \quad 0 \le x < x_*, \qquad w(x,t) = 0 \quad x > x_* . \tag{3.7}$$

The Nordgren problem consists in finding the solution of PDE (3.3) under the zero-opening initial condition (3.4) and the BC (3.5), (3.6). The solution should comply with (3.7).

*3.2. Speed equation, self-similar problem formulation, clear evidence that the problem is ill-posed*

Nordgren used the conditions (3.7) to find the front propagation, rather than the global mass balance commonly employed for this purpose (e. g., Howard & Fast 1970; Spence & Sharp 1985; Adachi & Detournay 2002; Savitski & Detournay 2002; Jamamoto et al. 2004; Garagash 2006; Adachi et al. 2007; Mitchel et al. 2007; Hu & Garagash 2010; Garagash et al. 2011). For our purposes, we employ the SE (2.12). In the case considered, it becomes:

$$v_* = -\frac{1}{3} k_l k_r \left. \frac{\partial w^3}{\partial x} \right|_{x=x_*} , \tag{3.8}$$

where according to (2.3) $v_* = dx_* / dt$ .

Introduce dimensionless variables:

$$x_d = \frac{x}{x_n}, \ x_{*d} = \frac{x_*}{x_n}, t_d = \frac{t}{t_n}, v_{*d} = \frac{dx_{*d}}{dt_d} = \frac{v_*}{v_{*n}}, w_d = \frac{w}{w_n}, p_d = \frac{p}{p_n}, q_d = \frac{q}{q_n}, q_{0d} = \frac{q_0}{q_n}$$

where $x_n = (k_l k_r)^{1/5} q_n^{3/5} t_n^{3/5}$, $v_{*n} = x_n / t_n$, $w_n = q_n t_n / x_n$, $p_n = 4k_r w_n$, and $t_n$, $q_n$ are arbitrary scales of the time and flux respectively. In terms of dimensionless values, the equations (3.1), (3.2) read $q_d = -w_d^3 \dfrac{\partial p_d}{\partial x_d}$, $p_d = 4w_d$; hence, $q_d = -\dfrac{\partial w_d^4}{\partial x_d}$. Then the PDE (3.3) and the BC (3.5), (3.6) become, respectively,

$$\frac{\partial^2 w^4}{\partial x^2} - \frac{\partial w}{\partial t} = 0, \tag{3.9}$$

$$-\frac{\partial w^4}{\partial x}\bigg|_{x=0} = q_0, \tag{3.10}$$

$$w(x_*, t) = 0. \tag{3.11}$$

From this point on, we omit the subscript $d$ at variables and consider only dimensionless values. The homogeneous conditions (3.4) and (3.7) do not change their form. Nordgren (1972) solved the problem by finite differences not using the SE.

In dimensionless variables, the SE (3.8) takes the form:

$$v_* = -\frac{4}{3} \frac{\partial w^3}{\partial x}\bigg|_{x=x_*}. \tag{3.12}$$

In compliance with the said in Sec. 2, the SE (3.12) gives a key to proper choice of the unknown function. Indeed, to have the front velocity finite, the partial derivative $\partial w^3 / \partial x$ should not be singular at the liquid front. This yields that $w^3$ is an analytical function of $x$ and it may be represented by a power series in $x_* - x$ at any instant $t$: $w^3(x,t) = \sum\limits_{j=0}^{\infty} a_j(t)(x_* - x)^j$. The zero-opening condition (3.11) gives $a_0(t) = 0$, while the SE (3.12) gives $a_1(t) = 0.75 v_*(t)$. Then at the vicinity of the crack tip we have $\qquad w^3(x,t) = a_1(t)(x_* - x) + O\big((x_* - x)^2\big)$ what means that the opening behaves as $w(x,t) = a_1^{1/3}(x_* - x)^{1/3} + O\big((x_* - x)^{2/3}\big)$. Therefore, near the crack tip, in contrast with $w^3$, the derivative $\partial w / \partial x$ of $w$ is singular as $(x_* - x)^{-2/3}$.

We see that using $w^3$ avoids unfavorable asymptotic behavior of $w$, while the SE (3.12) governs the linear asymptotic behavior of $w^3$. Hence, it is reasonable to use $w^3$ as the unknown function, rather than $w$ as used by Nordgren (1972) or $w^4$: in contrast with the derivatives of $w^3$, the derivatives $\dfrac{\partial w}{\partial x}$ and $\dfrac{\partial^2 w^4}{\partial x^2} = \dfrac{4}{3}\left(\dfrac{\partial w}{\partial x}\dfrac{\partial w^3}{\partial x} + w\dfrac{\partial^2 w^3}{\partial x^2}\right)$ are singular at the front $x = x_*$. In terms of $w^3$ PDE (3.9) becomes:

$$\frac{\partial^2 w^3}{\partial x^2} + \frac{1}{3w^3}\left(\frac{\partial w^3}{\partial x}\right)^2 - \frac{1}{4w^3}\frac{\partial w^3}{\partial t} = 0. \tag{3.13}$$

For further discussion we use the fact that the problem is self-similar which serves to reduce PDF (3.13) to an ordinary differential equation (ODE). We express the variables $x$ and $w$ via automodel variables $\xi$ and $\psi$ as $x = \xi t^{4/5}$, $w(x) = t^{1/5}\psi(xt^{-4/5})$. Then (3.13) becomes ODE with the unknown function $y(\xi) = \psi^3(\xi)$:

$$\frac{d^2 y}{d\xi^2} + a(y, dy/d\xi, \xi)\frac{dy}{d\xi} - \frac{3}{20} = 0, \tag{3.14}$$

where $a(y, dy/d\xi, \xi) = (dy/d\xi + 0.6\xi)/(3y)$. The BC (3.10) and (3.11) read:

$$\frac{dy}{d\xi}\bigg|_{\xi=0} = -0.75\frac{q_0}{\sqrt[3]{y(0)}}, \tag{3.15}$$

$$y(\xi_*) = 0, \tag{3.16}$$

and the SE (3.12) becomes:

$$\frac{dy}{\partial\xi}\bigg|_{\xi=\xi_*} = -0.6\xi_*. \tag{3.17}$$

Re-write (3.14) by using the expression for $a(y, dy/d\xi, \xi)$ as

$$y\frac{d^2y}{d\xi^2} + \frac{1}{3}\left(\frac{dy}{\partial\xi} + 0.6\xi\right)\frac{dy}{\partial\xi} - \frac{3}{20}y = 0, \tag{3.18}$$

In limit $\xi \to \xi_*$, for a solution, satisfying the BC (3.16), ODE (3.18) turns into the SE (3.17). Hence, for ODE (3.14), at the point $\xi = \xi_*$, we have imposed not only the BC (3.16) for unknown function $y$, but also the BC (3.17) for its derivative $dy/d\xi$. Note that equations (3.16), (3.17) imply that the factor $a$ in (3.14) is *finite* at the liquid front: $\lim_{\xi\to\xi_*} a(y, dy/d\xi, \xi) = -1/(3\xi_*)$.

It is easy to check by direct substitution that if $y_1(\xi_1)$ is the solution of the problem (3.14)-(3.17) for $q_0 = q_{01}$ with $\xi_{*} = \xi_{*1}$, then $y_2(\xi_2) = y_1(\xi_2\sqrt{k})/k$ is the solution of the problem (3.14)-(3.17) for $q_{02} = k^{-5/6}q_{01}$ with $\xi_{*2} = \xi_{*1}/\sqrt{k}$; herein, $k$ is an arbitrary positive number. This implies that $C_* = (q_0)^{0.6}/\xi_*$ and $C_0 = y(0)/\xi_*^2$ are constants not dependent on the prescribed influx $q_0$. As $\xi_* = (q_0)^{0.6}/C_*$, it is a matter of convenience to prescribe $q_0$ or $\xi_*$. A particular value of $q_0$ or $\xi_*$ may be also taken as convenient. Indeed, with the solution $y_1(\xi_1)$ for $q_0 = q_{01}$, we find the solution for any $q_0$: $y(\xi) = y_1(\xi\sqrt{k})/k$, where $k = (q_{01}/q_0)^{6/5}$, $\xi = \xi_1/\sqrt{k}$ ($\xi_* = \xi_{*1}/\sqrt{k}$).

Let us fix $\xi_*$. According to (3.16), (3.17), at the point $\xi_*$, we have prescribed both the function $y$ and its derivative $dy/d\xi$. Thus, for the ODE of the second order (3.14) we have a Cauchy (initial value) problem. Naturally, its solution defines $y(0)$ and $dy/d\xi|_{\xi=0}$ and consequently the flux $q_0$ at $\xi = 0$. Hence, even a small error when prescribing $q_0$ in (3.15), excludes the existence of the solution of the BV problem (3.14)-(3.16). Therefore, by Hadamard (1902) definition (see also Tychonoff 1963; Lavrent'ev & Savel'ev 1999), the BV problem (3.14)-(3.16) is ill-posed. It cannot be solved without a proper regularization (Tychonoff 1963; Lavrent'ev & Savel'ev 1999). To make conclusions on the accuracy of numerical results obtained without and with regularization, it is reasonable to obtain benchmarks.

## 4. Benchmark solution

The initial value (Cauchy) problem (3.14), (3.16), (3.17) is well-posed. Thus its solution provides the needed benchmarks. To solve the system we transformed the problem (3.14), (3.16), (3.17) to the equivalent problem in two unknowns $Y_1(\xi) = y(\xi)$ and $Y_2(\xi) = dy/d\xi$. This yields the system of two ODE:

$$\frac{dY_1}{d\xi} = Y_2$$

$$\frac{dY_2}{d\xi} = -a(Y_1, Y_2, \xi)Y_2 + \frac{3}{20} \tag{4.1}$$

under the Cauchy conditions at the point $\xi_*$, corresponding to (3.16) and (3.17),

$$Y_1(\xi_*) = 0, \quad Y_2(\xi_*) = -0.6\xi_*. \tag{4.2}$$

The Cauchy problem (4.1), (4.2) is solved by using the fourth order Runge-Kutta scheme (see, e.g. Epperson 2002). Calculations were preformed with double precision. For certainty, we set $\xi_* = \xi_{*1} = 1$ (calculation with other values of $\xi_*$ gave the same results for $C_*$ and $C_0$ to the seventh significant digit including). The integration step $\Delta\xi$ was changed from $10^{-2}$ to $10^{-5}$, and the number of steps was consequently changed from $10^2$ to $10^5$ to reach the inlet point $\xi = 0$. The number of iterations for the nonlinear factor $a(y, dy/d\xi, \xi)$ was changed from 20 to 1000. We could see that the step $10^{-5}$ and the number of iterations 50 are sufficient to guarantee at least six correct digits. The values of the constants $C_*$ and $C_0$ are evaluated to the accuracy of seven digits:

$$C_* = 0.7570913, \quad C_0 = 0.5820636.$$

For values depending on $\xi_*$, we shall use the subscript 1 when they correspond to $\xi_* = \xi_{*1} = 1$. Thus we have:

$$q_{01} = C_*^{5/3} = 0.6288984, \quad \psi_1(0) = \sqrt[3]{C_0} = 0.8349418.$$

Values for an arbitrary flux $q_0$ may be obtained as

$$\xi_* = q_0^{0.6} / C_* = 1.320844 q_0^{0.6}, \quad \psi(0) = \sqrt[3]{C_0\xi_*^2} = 1.0051356 q_0^{0.4}. \tag{4.3}$$

For the value $q_0 = 2/\pi$, used by Nordgren (1972), equations (4.3) give $\xi_* = 1.0073486$, $\psi(0) = 0.8390285$ against the values given by this author to the accuracy of about one percent: $\xi_* = 1.01$, $\psi(0) = 0.83$. The values of $\psi_1 = \sqrt[3]{y_1}$ and $d\psi_1^3 / d\xi_1$ are presented in Table 1 with five correct digits.

Values of $\psi(\xi)$ and $d\psi^3 / d\xi$ for an arbitrary prescribed flux $q_0$ may be obtained from those in the Table 1 as $\psi = \sqrt[3]{y_1(\xi\sqrt{k})/k}$, $d\psi^3 / d\xi = (d\psi_1^3 / d\xi_1)/\sqrt{k}\big|_{\xi_1 = \xi\sqrt{k}}$ with $k = (q_{01}/q_0)^{1.2} = 0.5731872 / q_0^{1.2}$.

*Comment 3.* Table 1 shows that the derivative $d\psi^3 / d\xi$, defining the particle velocity, is nearly constant along the entire liquid being close to its limit value at the front $\xi = \xi_*$. This implies that $\psi^3$ is approximately linear in $\xi$. Next, from the BC (3.16) we obtain $\psi^3(\xi)/\psi^3(0) \approx 1 - \xi/\xi_*$. Hence, the approximate analytical solution of the Nordgren problem is given by the equation:

$$w(x,t)/w(0,t) \approx (1 - x/x_*)^{1/3} \tag{4.4}$$

with $w(0,t) = t^{0.2}\psi(0)$ and $x_*(t) = \xi_* t^{0.8}$; for a given flux $q_0$, the values of $\psi(0)$ and $\xi_*$ are found from (4.3). As clear from Table 1, the error of (4.4) does not exceed one percent. The graph corresponding to the approximate solution (4.4) is indistinguishable from that given by Nordgren (1972). Naturally, the asymptotic of the solution (4.4) agrees with the predicted asymptotic behavior $w(x,t) \approx a_1^{1/3}(x_* - x)^{1/3}$ near the crack tip because the conditions (4.2) guarantee that $\psi^3$ is proportional to $\xi_* - \xi$, and consequently, $w^3$ is proportional to $x_* - x$.

The accurate data $\xi_* = 1.320844 q_0^{0.6}$, $\psi(0) = 1.0051356 q_0^{0.4}$, $\psi(\xi)$ and $d\psi^3 / d\xi = (d\psi_1^3 / d\xi_1)/\sqrt{k}\big|_{\xi_1 = \xi\sqrt{k}}$ with $k = 0.5731872 / q_0^{1.2}$ serve to estimate errors when solving the Nordgren problem as a BV problem. When $\xi_*$, $\psi(\xi)$ and $d\psi^3 / d\xi$ are known, we can find the front location $x_*(t)$, the front velocity $v_*(t)$, the opening $w(x,t)$ and the particle velocity $v(x,t)$ as $x_*(t) = \xi_* t^{0.8}$, $v_*(t) = 0.8\xi_* t^{-0.2}$, $w(x,t) = t^{0.2}\psi(xt^{-0.8})$ and $v(x,t) = -\dfrac{4}{3} t^{0.2} \dfrac{d\psi^3}{d\xi}\Big|_{\xi = xt^{-0.8}}$.

## 5. Straightforward solving self-similar BV problem. Method of regularization

### 5.1. Straightforward integration by finite differences

Forget for a while about the SE and all the said on its influence on a BV problem. Let us see what happens when solving the BV problem (3.14)-(3.16) in a common way by finite differences.

We performed hundreds of numerical experiments with various numbers of nodal points and iterations and different values of the prescribed influx $q_{01}$ at the inlet. Finite difference approximations of second order for $d^2y/d\xi^2$ and $dy/d\xi$ were combined with iterations for $a(y, dy/d\xi, \xi)$. Up to 100 000 nodal points and up to 1500 iterations were used in attempts to reach the accuracy of three correct digits, at least. The attempts failed: by no means could we have more than two correct digits. Moreover, the results always strongly deteriorate near the liquid front. *The numerical results clearly demonstrate that the BV problem (3.14)-(3.16) is ill-posed. It cannot be solved accurately without regularization.*

As illustration, the dashed line in Fig. 2 presents a typical graph of $d\psi^3/d\xi$, obtained under the Nordgren boundary value $q_{01} = 2/\pi$. For comparison, the benchmark values of $d\psi^3/d\xi$, calculated by using the Table 1, are shown by the solid line with markers. Obviously, the results strongly deteriorate near the liquid front $\xi = \xi_*$ (in the considered example, the benchmark value of $\xi_*$ equals 1.0073486).

*Comment 4.* Using the variable $\psi^3(\xi)$, which is linear near the liquid front, removes a suggestion that the deterioration is caused by singularity of $d\psi/d\xi$ and $d^2\psi/d\xi^2$ at the point $\xi = \xi_*$.

*Comment 5.* It is worth noting that the accuracy of *two* correct digits was obtained at points not too close to the front even when using a rough mesh with a hundred or even only ten nodes. This indicates that using a rough mesh may serve to regularize a problem when high accuracy is not needed.

### *5.2. ε - regularization.*

The numerical experiments evidently confirm that the considered ill-posed BV problem (3.14)-(3.16) cannot be solved accurately without regularization. A regularization method is suggested by the conditions (3.16), (3.17). Indeed, we may use them together to get the approximate equation $y \approx 0.6\xi_*(\xi_* - \xi)$ near the front. Hence, instead of prescribing the BC (3.16) at the liquid front $\xi = \xi_*$, where it is implicitly complimented by the SE (3.17), we may impose the boundary condition, which combines (3.16) and (3.17) at a point $\xi_\varepsilon = \xi_*(1 - \varepsilon)$ at a small relative distance $\varepsilon = 1 - \xi_\varepsilon/\xi_*$ from the front:

$$y(\xi_\varepsilon) = 0.6\xi_*^2\varepsilon . \tag{5.1}$$

The BV problem (3.14), (3.15), (5.1) is well-posed and may be solved by finite differences. Numerical implementation of this approach shows that with $\varepsilon = 10^{-3}$, $10^{-4}$ the results for the step $\Delta\varsigma = \Delta\xi/\xi_* = 10^{-3}$, $10^{-4}$, $10^{-5}$, $10^{-6}$ coincide with those of the benchmark solution. The time expense is fractions of a second. The results are stable if ε and $\Delta\varsigma$ are not simultaneously too small (both ε and $\Delta\varsigma$ are greater than $10^{-5}$).

As could be expected, the results deteriorate when both the regularization parameter ε and the step $\Delta\varsigma$ become too small. Specifically, when $\varepsilon = \Delta\varsigma = 10^{-6}$, the results are completely wrong. Actually, in this case, to the accuracy of computer arithmetic, the problem is solved without regularization.

We could also see that with growing step $\Delta\varsigma$, the accuracy decreases and for a coarse mesh it actually does not depend on the regularization parameter. In particular, for a quite coarse mesh with the step $\Delta\varsigma = 0.1$, the accuracy is about one percent, and the results stay the same to this accuracy for any $\varepsilon$ from $10^{-2}$ to $10^{-9}$.

The essence of the suggested regularization consists in using the SE together with a prescribed BC to formulate a BC at a small distance behind the liquid front rather than on the front itself. We call such an approach ε-regularization. The next section contains its extension to the cases when a self-similar formulation is not available or is not used.

## 6. Straightforward solution of starting BV problem. Regularization

### 6.1. Straightforward integration by time steps with finite differences on a time step

Forget again about the SE and its influence on numerical solution of a BV problem. Try to solve the starting Nordgren problem by common finite differences. Nordgren (1972) used straightforward numerical integration of the problem (3.9)-(3.11) under the zero-opening initial condition with the conditions that opening is positive behind the liquid front and zero ahead of it. This author applied Crank-Nicolson finite difference scheme to approximate PDF (3.9) and to meet the BC (3.10), (3.11).

The resulting non-linear tridiagonal system was linearized by employing linear approximation of $w^4$. Nordgren (1972) does not include details of calculations on the initialization, the time step, the number of nodes in spatial discretization, the number of iterations, stability of numerical results and expected accuracy. To obtain knowledge on these issues, we also solved the system (3.9)-(3.11) in a straightforward way by using the Crank-Nicolson scheme. The results are as follows.

Actually performing 20 iterations to account for the non-linear term $w^4$ is sufficient to reproduce four digits of the fracture opening, except for close vicinity of the liquid front (Increasing the number to 100 iterations does not improve the solution for all tested time and spatial steps.). For various time steps ($\Delta t = 10^{-2}$, $10^{-3}$, $10^{-4}$) and different spatial steps ($\Delta x = 10^{-2}$, $10^{-3}$, $10^{-4}$) taken in various combinations, the results are stable along the main part of the interval $[0, x_*(t)]$, but deteriorate and are unreliable in close vicinity of the front ($1 - x/x_*(t) < 0.001$). This yields changes in the third digit of $\xi_*$ and $\psi(0)$, calculated by using $x_*(t)$ and $w(0,t)$. The asymptotic behavior (4.4), as $x \to x_*(t)$, is reproduced near the front except for its close vicinity. The results coincide with those given by Nordgren (1972) to the accuracy of two significant digits accepted in his work.

In all the calculations, by no means could we have a correct third digit. Similar to self-similar solution, fine meshes did not improve the accuracy as compared with a rough mesh having the step $\Delta\varsigma = \Delta x / x_* = 0.01$. The results clearly show that the problem cannot be solved accurately without regularization.

*Comment 6.* As mentioned in Sec. 3, the variable $w$ has a singular partial derivative $\partial w / \partial x$ at the liquid front. To remove the influence of the singularity, we also solved the problem by using $w^3$ as an unknown function, because according to the SE (3.12) its spatial derivative is not singular. The conclusions when using $w^3$ are the same as those above. Again, by no means could we have reliable a third digit, and results strongly deteriorated at a close vicinity of the liquid front. This shows that the inaccuracy is caused not by singularity of the derivative $\partial w / \partial x$ at the liquid front. It is caused by the fact the BV problem, when solved by common finite differences, appears ill-posed.

### 6.2. Reformulation of PDE to form appropriate for using ε - regularization. Numerical results

Extension of ε - regularization to solve PDE requires the combined use of the BC (3.11) on the front with the SE (3.12) to impose a BC at a small relative distance ε from the front. The distance being relative, we need to count it in the local system with the origin at the front. Hence it is reasonable to introduce the relative distance $\eta = (x_* - x)/x_*$ from the front. The relative distance from the inlet is $\varsigma = 1 - \eta = x/x_*(t)$. The partial time derivative $\partial\varphi/\partial t|_{x=const}$ of a function $\varphi(x,t)$, which enters PDE and is evaluated under constant $x$, should be transformed into the partial time derivative $\partial\Phi/\partial t|_{\varsigma=const}$ of the function $\Phi(\varsigma,t) = \varphi(\varsigma x_*(t),t)$ evaluated under constant $\varsigma$. Omitting routine details of the change of variables, we have the transformation:

$$\left.\frac{\partial\varphi}{\partial t}\right|_{x=const} = \left.\frac{\partial\Phi}{\partial t}\right|_{\varsigma=const} - \varsigma\frac{v_*(t)}{x_*(t)}\frac{\partial\Phi}{\partial\varsigma}, \qquad (6.1)$$

where $v_*(t) = dx_* / dt$. When using the variable $\varsigma$ and transformation (6.1), equation (3.13) becomes:

$$\frac{\partial^2 Y}{\partial \varsigma^2} + A(Y, \partial Y / \partial \varsigma, x_* v_* \varsigma)\frac{\partial Y}{\partial \varsigma} - B(Y, x_*)\frac{\partial Y}{\partial t} = 0, \tag{6.2}$$

where $Y(\varsigma, t) = w^3(\varsigma x_*(t), t)$, $A(Y, \partial Y / \partial \varsigma, x_* v_* \varsigma) = \dfrac{\partial Y / \partial \varsigma + 0.75 x_* v_* \varsigma}{3Y}$, $B(Y, x_*) = \dfrac{x_*^2}{4Y}$.

The BC (3.10), (3.11) in new variables read:

$$-\frac{4}{3}\frac{\sqrt[3]{Y}}{x_*}\frac{\partial Y}{\partial \varsigma}\bigg|_{\varsigma=0} = q_0, \tag{6.3}$$

$$Y(\varsigma, t)\big|_{\varsigma=1} = 0. \tag{6.4}$$

The SE (3.12) takes the form:

$$\frac{\partial Y}{\partial \varsigma}\bigg|_{\varsigma=1} = -0.75 x_* v_*. \tag{6.5}$$

Prove firstly, that the problem (6.2)-(6.4) is ill posed because like the self-similar formulation, the SE (6.5) is met identically by a solution of PDF (6.2) under the BC (6.4) of zero opening at the liquid front. Indeed, re-write (6.2) by using the expression for $A(Y, \partial Y / \partial \varsigma, x_* v_* \varsigma)$ as

$$Y\frac{\partial^2 Y}{\partial \varsigma^2} + \frac{1}{3}\left(\frac{\partial Y}{\partial \varsigma} + 0.75 x_* v_* \varsigma\right)\frac{\partial Y}{\partial \varsigma} - 0.75 x_*^2 \frac{\partial Y}{\partial t} = 0. \tag{6.6}$$

In limit $\varsigma \to 1$ ($x \to x_*$), for a solution, satisfying the BC (6.4), PDE (6.6) turns into the SE (6.5). Hence, for PDE (6.2), at the point $\varsigma = 1$, we have imposed not only the BC (6.4) for unknown function $Y$, but also the BC (6.5) for its spatial derivative $dY / d\varsigma$. Therefore, we have two rather than one BC at $\varsigma = 1$ and the problem appears ill-posed. Consequently, the starting problem (3.9)-(3.11) is ill-posed, as well, what explains the failure to solve it to the accuracy greater than two correct digits. Note that equations (6.4), (6.5) imply that the factor $A(Y, \partial Y / \partial \varsigma, x_* v_* \varsigma)$ in (6.2) is finite at the liquid front despite its denominator $3Y$ turns to zero: $\lim\limits_{\xi \to \xi_*} A(Y, dY / d\varsigma, x_* v_* \varsigma) = -1/3$.

The regularization of the problem (6.2)-(6.4) follows the line used for the self-similar formulation. Like the self-similar formulation, the BC (6.4) and the SE (6.5) yield the approximate equation near the liquid front $\varsigma = 1$:

$$Y(\varsigma, t) \approx 0.75 x_*(t) v_*(t)(1 - \varsigma), \tag{6.6}$$

which defines the asymptotic behavior of the solution when $\varsigma \to 1$. As mentioned, the non-linear multiplier $A(Y, \partial Y / \partial \varsigma, x_* v_* \varsigma)$ in PDF (6.2) is non-singular at the liquid front $\varsigma = 1$. The factor $B(Y, x_*)$ is finite except for the liquid front ($\varsigma = 1$), because the opening is positive behind the front. Hence, similar to (5.1), we may impose the BC at the relative distance $\varepsilon$ from the liquid front:

$$Y(\varsigma_\varepsilon, t) = 0.75 x_*(t) v_*(t)\varepsilon, \tag{6.7}$$

where $\varsigma_\varepsilon = 1 - \varepsilon$. In contrast with the problem (6.2)-(6.4), the problem (6.2), (6.3), (6.7) does not involve an additional BC. We may expect that it is well-posed and provides the needed regularization. Extensive numerical tests confirm the expectation.

We solved the problem (6.2), (6.3), (6.7) by using the Crank-Nicolson scheme and iterations for non-linear multipliers $A(Y, \partial Y / \partial \varsigma, x_* v_* \varsigma)$, $B(Y, x_*)$ at a time step. The velocity $v_*(t)$ is also iterated by using the equation following from (6.6):

$$\frac{\partial Y}{\partial \varsigma}\bigg|_{\varsigma=\varsigma_\varepsilon} = -\frac{3}{4} x_* v_*. \tag{6.8}$$

The condition (6.8) expresses (with an accepted tolerance) the continuity of the particle velocity at the point $\varsigma = \varsigma_\varepsilon$. After completing iterations, the final value of $v_*(t)$ shows the new coordinate of the liquid front $x_*(t + \Delta t) = x_*(t) + v_*(t)\Delta t$. The value $x_*(t + \Delta t)$ is used on the next time step.

*Initialization.* In the considered problem there is no characteristic time and length. Thus, the starting equation $x_*(0) = 0$ should be adjusted to this uncertainty. The adjustment concerns the initialization of time stepping. At $t = 0$ the liquid front coincides with the inlet. The opening is also zero. Then for 'small' time, to have the factor $A(Y, \partial Y/\partial \varsigma, x_* v_* \varsigma)$ finite in the entire liquid, it should be $\partial Y/\partial \varsigma = x_* v_* f'(\varsigma)$ with $f'(\varsigma) \to -0.75$ as $\varsigma \to 1$. Hence for 'small' time, we have $Y = x_* v_* f(\varsigma)$ with $f(1) = 0$. Substitution of $Y$ and $\partial Y/\partial \varsigma$ into the BC (6.3), using $v_* = dx_*/dt$ and integration of the resulting ODE yield $x_*(t) = Ct^{0.8}$, where $C = 1.25^{0.8}\left(-\dfrac{0.75 q_0}{\sqrt[3]{f(0)f'(0)}}\right)^{0.6}$ is a constant. Then $v_*(t) = 0.8 Ct^{-0.2}$ and $Y = 0.8 C^2 t^{0.6} f(\varsigma)$. Insertion of these equations into PDF (6.2) gives ODE (3.14) with $y = 0.8 f$ and $\varsigma = \xi/\xi_*$. Similarly, the BC (6.3), (6.4) and the SE (6.5) turn into the corresponding equations (3.15), (3.16) and (3.17). The constant $C$ actually represents $\xi_*$, corresponding to the prescribed flux $q_0$.

We see that when using time stepping, initialization requires solving the self-similar problem. The latter being ill-posed, the solution is obtained by $\varepsilon$ - regularization as explained in Sec. 5.2. Having its solution, the initial data for an arbitrary chosen initialization time $t_0$ are found as $x_*(t_0) = \xi_* t_0^{0.8}$, $v_*(t_0) = 0.8 \xi_* t_0^{-0.2}$, $Y(\varsigma, t_0) = \xi_*^2 t_0^{0.6} \psi(\varsigma \xi_*)$. For certainty, in the following calculations we set $q_0 = 1$, $t_0 = 0.01$. The same regularization parameter $\varepsilon$ and the same spatial step $\Delta \varsigma$ were used for both the initialization and for each of the time steps.

*Numerical results.* Two objectives were sought in numerical tests. First, we wanted to check the efficiency of $\varepsilon$-regularization, that is, its accuracy, stability and robustness. To this end, we used small relative distance $\varepsilon$, small spatial step $\Delta \varsigma$, small time step $\Delta t$, and a large number of time steps. Secondly, we checked if the beneficial features of coarse meshes, observed for self-similar formulation, hold in time steps. Exploratory calculations have shown that fifty iterations in non-linear terms at a time step are sufficient to reproduce seven significant digits. Thus, in further tests the number of iterations was set equal to fifty. The benchmark solution served to evaluate the accuracy. In particular, for $q_0 = 1$, the benchmark values of the front position and the front velocity are $x_*(t) = 1.3208446 t^{0.8}$ and $v_*(t) = 1.0566757 t^{-0.2}$, respectively.

*Results for fine meshes and small time steps.* Obviously, the accuracy at the first time steps depends on the accuracy of the data obtained at the initialization stage. We could see that $\varepsilon = 0.0001$ and $\Delta \varsigma = 0.01$ provide the accuracy of 0.026% for $\xi_*$. To keep the accuracy on this level for $x_*(t)$ and $v_*(t)$ at first time steps, the step $\Delta t$ should be notably less than $t_0$. By taking $\Delta t = 0.01 t_0 = 10^{-4}$, we had the accuracy of some 0.03% for both $x_*(t)$ and $v_*(t)$ at first hundred steps. With further growth of the time, the accuracy of $v_*(t)$ stayed on the same level, while the relative error of $x_*(t)$ decreased. In particular, at the step $m = 1000$ ($t = 0.11$), it is 0.0077%; at the step $m = 20000$ ($t = 2.01$) it is 0.0043%. The decrease of the relative error in $x_*$ for large time is related to the growth of the absolute value of $x_*$ (Recall that $x_*(t) = \xi_* t^{0.8}$.). The time expense on a conventional laptop for 20000 steps with 50 iterations at a step is near 15 s. Note that in view of the growth of $x_*(t)$ in time, there is no need to have the time step constant. By using exponentially growing time steps, the number of time steps may be drastically reduced without loss of the accuracy. For instance, with the same time expense we could reach time $t = 36128$ ($x_* = 5889.4$) with no loss of the accuracy. The number of iterations could be reduced as well.

There were no signs of instability in these and many other specially designed experiments. We could see that for $10^{-2} > \varepsilon > 10^{-4}$ and $\Delta \varsigma \leq 0.01$, the results are accurate and stable in a wide range of values of the time step and for very large number (up to 100000) of steps. Therefore $\varepsilon$ - regularization (6.7) is quite efficient.

*Results for coarse mesh and large number of time steps.* For the coarse mesh $\Delta\varsigma = 0.0(9)$ and $\varepsilon = 0.1$ at the initialization stage, we have one percent accuracy except for close vicinity of the liquid front. There is no need to take time steps notably less than the initial time. When taking $\Delta t = 0.25 t_0 = 0.0025$, we had $x_*$ with the error not exceeding 1.2% at any time instant for 20000 equal steps. Again, the relative error decreased with growing $x_*$. Using exponentially growing time step serves to obtain $x_*(t)$ and $v_*(t)$ with the accuracy of one percent even when $x_*(t)$ becomes of order 10000 (recall that the starting value is $x_*(t_0) = 0.033170$). As the number of nodes is small (only 11), the time expense is fractions of a second. Again, there were no signs of instability. Moreover, increasing the initial time step 3-10-fold, influenced the accuracy only of the first steps, while with growing time, the accuracy kept to the mentioned level of about 1%. Hence, use of a coarse mesh efficiently maintains this accuracy.

## 7. Extension of the regularization method to 2D fracture propagation

According to the rationale presented in the preceding section, it appears that the strategy of using $\varepsilon$ - regularization when tracing 2-D hydrofracture propagation is as follows. At each point of the liquid front we introduce the local coordinate system moving with the front in the direction normal to the front. We transform field equations to this system. The prescribed BC at the front and the SE associated with front points are transformed accordingly. They are combined to formulate a new BC at a small distance behind the front. Below, we follow this path.

We introduce the local Cartesian coordinates $x_1'O_1'x_2'$ with the origin $O_1'$ at a point $\mathbf{x}_*$ at the liquid front and the axis $x_1'$ opposite to the direction of external normal to the front at this point. As the flux $q_*$ at the front is co-linear to the normal, its tangential component is zero. For the point $\mathbf{x}_*$, equation (2.11) written in the system $x_1'O_1'x_2'$ becomes a scalar equation $q_{n*} = D(w, p)\dfrac{\partial p}{\partial x_1'}\bigg|_{x_1'=0}$, and the SE (2.12) reads;

$$v_* = \frac{1}{w_*} D(w, p)\frac{\partial p}{\partial x_1'}\bigg|_{x_1'=0}, \qquad (7.1)$$

where $v_* = dx_{n*}/dt$ is the absolute value of the front velocity. Combining (7.1) with the BC (2.16) we obtain:

$$\int_{p_0}^{p} \frac{1}{w} D(w, p)dp \approx v_* x_1'. \qquad (7.2)$$

Equation (7.2) serves to impose the BC at a small distance $x_1' = \eta_\varepsilon$ from the front:

$$\int_{p_0}^{p_\varepsilon} \frac{1}{w} D(w, p)dp = v_* \eta_\varepsilon, \qquad (7.3)$$

where $p_\varepsilon = p(\eta_\varepsilon)$ is the pressure at the point $\eta_\varepsilon$. The $\varepsilon$ - regularization consists in using the BC (7.3) instead of the BC (2.16).

In numerical calculations with $\varepsilon$ - regularization, it is reasonable to use the aforementioned advantages of variables which express the particle velocity $\mathbf{v} = \mathbf{q}/w$. In the local system $x_1'O_1'x_2'$ moving with the front, for points close to the front, the normal component of the velocity is: $v = \dfrac{1}{w} D(w, p)\dfrac{\partial p}{\partial x_1'}$. Evaluating the partial time derivative under constant $x_1'$ rather than at constant global coordinate, the PDF (2.9) at points near the front takes the form (2.17). As noted in Sec. 2, it is reasonable to transform PDF (2.17) to (2.18) by introducing the variable $y = w^{1/\alpha}$, where the exponent $\alpha$ accounts for asymptotic behavior of the opening at the front.

As $\eta_\varepsilon$ is small, the velocity $v(\eta_\varepsilon)$ is close to $v_*$. This serves to employ the equation

$$v(\eta_\varepsilon) = v_* \tag{7.4}$$

for iterations in $v_*$ at a time step. For the Nordgren problem, $\alpha = 1/3$, $p = 4w$, $p_0 = 4w(x_*) = 0$, $\eta_\varepsilon = \varepsilon x_*$; then equations (2.18), (7.3) and (7.4) are easily transformed to (6.2), (6.7) and (6.8), respectively.

## 8. Conclusions

The paper presents the following conclusions.

(i) The *speed function* for fluid flow in a thin channel is given by the ratio of the flux through a cross section to the channel width at the front. Its specification for hydraulic fracturing follows from the equation of Poiseuille type connecting the flux with gradient of pressure. The speed function, as the basis of the theory of propagating interfaces, facilitates employing such methods as level set methods and fast marching methods. The *speed equation* for hydraulic fracturing is a general condition at the liquid front not dependent on a particular BC defined by a physical situation ahead of the front, liquid properties, leak-off through the channel walls and/or presence of distributed sources. When there is no lag between the liquid front and the crack tip, both the flux and opening are zero at the liquid front; in this case, the SE is fulfilled in the limit when a point behind the front tends to the front.

(ii) Using the SE gives a key to proper choice of unknown functions when solving a hydraulic fracture problem numerically. Specifically, the particle velocity, averaged over the opening, is a good choice, because it is non-singular at the front and non-zero in entire flow region. Using the velocity as an unknown may also serve to avoid the stiffness of the system of differential equations obtained in a conventional way. Another proper unknown is the opening to the degree $1/\alpha$, where $\alpha$ is non-negative exponent, characterizing the asymptotic behavior of the opening near the front.

(iii) Existence of the SE also discloses the crucial feature of the problem: for zero or small lag, at the points of the front we actually have prescribed *two* rather than one BC; the BV problem appears ill-posed when solved numerically. It requires a proper regularization to have accurate and stable numerical results.

(iv) Self-similar formulation of the Nordgren problem, reducing PDE to ODE, unambiguously demonstrates that the problem is ill-posed when considered as a BV problem. Numerical experiments confirm this theoretical conclusion: by no means could we obtain more than two correct digits when solving the problem without regularization, and the results always strongly deteriorate near the liquid front.

(v) The solution of the self-similar problem, obtained when solving it as a Cauchy (initial value) problem by the Runge-Kutta method, provides benchmarks with at least five correct digits at entire liquid including its front. These numerical results may serve for testing methods of hydraulic fracture simulation.

(vi) Studying of the Nordgren problem, accounting for the SE, suggests a means to overcoming the difficulty by $\varepsilon$-regularization. It consists in employing the SE together with a prescribed BC on the front to formulate a new BC at a small distance behind the front rather than on the front itself. Application of $\varepsilon$-regularization and comparing the results with the benchmarks shows that it is robust and provides highly accurate and stable results.

(vii) Numerical experiments also disclose that using coarse spatial meshes is beneficial and serves as a specific regularization when high accuracy is not needed. Still, although involving analytical work, ε - regularization is superior in the possibility to guarantee accurate and stable results and to evaluate the accuracy of calculations; it is competitive in time expense.

(viii) The suggested ε - regularization may serve as a means to obtain accurate and stable results for simulation of hydrofracture with zero or small lag. When tracing 2-D hydrofracture propagation, ε - regularization is obtained by writing equations in the local coordinate system moving with the liquid front.

***Acknowledgment.*** The author appreciates the support of the EU Marie Curie IAPP program (Grant # 251475).

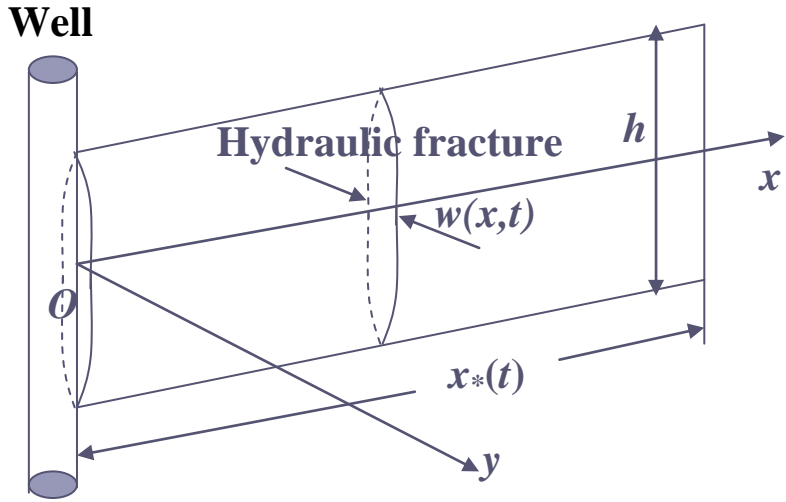

Fig. 1  Scheme of the problem on hydraulic fracture propagation

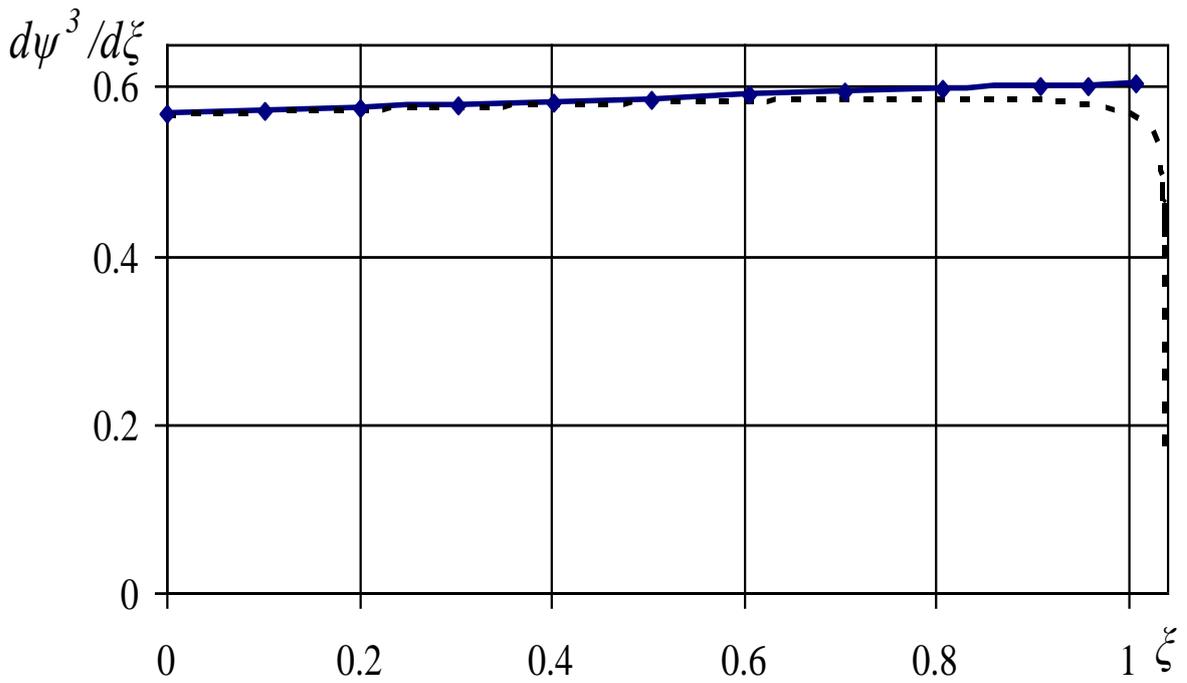

Fig. 2 Graphs of the normalized velocity $d\psi^3/d\xi$ in self-similar formulation of Nordgren's problem

———————  benchmark solution

- - - - -  solution of ill-posed boundary value problem